\documentclass{PoS}

\title{Dynamical spin effects in the pseudoscalar nonet within holographic QCD}

\ShortTitle{Dynamical spin effects in pseudoscalar mesons}

\author{Mohammad Ahmady\\
        Physics Department, Mount Allison University, New-Brunswick, E4L 1E6, Canada\\
       E-mail: \email{mahmady@mta.ca}}
\author{Chandan Mondal\\
   Institute of Modern Physics, Chinese Academy of Sciences, Lanzhou-730000, China\\
       E-mail: \email{mondal@impcas.ac.cn}}
\author{\speaker{Ruben Sandapen}\\
        Physics Department, Acadia University,
              Nova-Scotia, B4P 2R6, Canada\\\
        E-mail: \email{ruben.sandapen@acadiau.ca}}

\abstract{We investigate the importance of dynamical spin effects in the holographic light-front wavefunctions of the pseudoscalar mesons. We find that these effects are crucial to describe the pion data while they are not necessary to describe the available kaon data. For $\eta-\eta^\prime$ system, we find that dynamical spin effects are required to describe their transition form factors data.}

\FullConference{XIII Quark Confinement and the Hadron Spectrum - Confinement2018\\
		31 July - 6 August 2018\\
		Maynooth University, Ireland}
\usepackage{graphicx}
\usepackage{mathptmx}      
\usepackage{latexsym}
\usepackage{amssymb}
\usepackage{dsfont}
\usepackage{mathtools}
\usepackage{array,multirow}

\begin{document}
\section{Introduction}
\label{intro}

The holographic light-front Schr\"odinger Equation for mesons \cite{deTeramond:2005su,Brodsky:2006uqa,deTeramond:2008ht,Brodsky:2014yha}, 
 \begin{equation}
			\left(-\frac{\mathrm{d}^2}{\mathrm{d}\zeta^2}-\frac{1-4L^2}{4\zeta^2} + U_{\mathrm{eff}}(\zeta) \right) \phi(\zeta)=M^2 \phi(\zeta) \;,
	\label{hSE}
	\end{equation}
is derived within the semiclassical approximation of light-front QCD \cite{Brodsky:2014yha} where quarks are taken to be massless and quantum loops are neglected. The holographic variable\footnote{In this paper, $\bar{x}= 1-x$.} 
\begin{equation}
	\mathbf{\zeta}^2 = x\bar{x} \mathbf{b}^2_{\perp}
	\end{equation}
	is analogous to the radial variable in the ordinary Schr\"odinger Equation and it 
	maps onto the fifth dimension in anti-de Sitter (AdS) space so that Eq. \ref{hSE} also describes the propagation of weakly-coupled spin-$J$ modes in a modified AdS space. The confining QCD potential is determined by the form of the dilaton field which distorts the conformally invariant geometry of AdS space. Specifically, the potential is given by
\cite{Brodsky:2014yha}
	\begin{equation}
	U_{\mathrm{eff}}(\zeta)= \frac{1}{2} \varphi^{\prime \prime}(z) + \frac{1}{4} \varphi^{\prime}(z)^2 + \frac{2J-3}{2 z} \varphi^{\prime}(z) 
	\label{dilaton-potential}
	\end{equation}	
where $\varphi(z)$ is the dilaton field in AdS space. It turns out that the only way to introduce a mass scale in the holographic Schr\"odinger  without destroying the conformal invariance of the underlying semiclassical action, is via a harmonic oscillator potential i.e.  $U^{\mathrm{dAFF}}_{\mathrm{eff}}=\kappa^4 \zeta^2$ \cite{Brodsky:2013ar}. To recover this harmonic potential, the dilaton field has to be quadratic, i.e. $\varphi(z)=\kappa^2 z^2$ so that Eq. \ref{dilaton-potential} then implies that 
	\begin{equation}
		U_{\mathrm{eff}}(\zeta)=\kappa^4 \zeta^2 + 2 \kappa^2 (J-1)	
	\label{hUeff}	
	\end{equation}
	where $J=L+S$. Solving the holographic Schr\"odinger Equation with the confining potential given by Eq. \ref{hUeff} yields the mass spectrum
\begin{equation}
 	M^2= 4\kappa^2 \left(n+L +\frac{S}{2}\right)\;
 	\label{mass-Regge}
 \end{equation}
and the wavefunctions
 \begin{equation}
 	\phi_{nL}(\zeta)= \kappa^{1+L} \sqrt{\frac{2 n !}{(n+L)!}} \zeta^{1/2+L} \exp{\left(\frac{-\kappa^2 \zeta^2}{2}\right)} \nonumber \\ \times ~ L_n^L(\kappa^2 \zeta^2)\;.
 \label{phi-zeta}
 \end{equation}
The first non-trivial  prediction is that the lowest lying bound state, with quantum numbers $n=L=S=0$, is massless: $M^2=0$.  This state is naturally identified with the pion since the latter is expected to be massless in chiral limit $m_f \to 0$. Note that the harmonic oscillator confining potential uniquely leads to a massless pion \cite{Brodsky:2013npa}: if a more general potential $U_{\mathrm{eff}}(\zeta) \propto \zeta^p$ is considered, then $M_{\pi}=0$ only if $p=2$. The complete meson light-front wavefunction is given by \cite{Brodsky:2014yha}
\begin{equation}
	\Psi(x,\zeta,\varphi)=\frac{\phi(\zeta)}{\sqrt{2\pi \zeta}} X(x) e^{iL\varphi} \;,
	\end{equation}
where $X(x)=\sqrt{x\bar{x}}$ \cite{Brodsky:2008pf}. 

The normalized holographic light-front wavefunction for a ground state meson is then given by
 \begin{equation}
 	\Psi (x,\zeta^2) = \frac{\kappa}{\sqrt{\pi}} \sqrt{x \bar{x}}  \exp{ \left[ -{ \kappa^2 \zeta^2  \over 2} \right] } \;.
\label{pionhwf} 
\end{equation}
Going beyond the semiclassical approximation, one can account for 
non-vanishing quark masses. This is carried out in Ref. \cite{Brodsky:2014yha}, yielding
\begin{equation}
\Psi (x,\zeta^2) = \mathcal{N} \sqrt{x \bar{x}}  \exp{ \left[ -\frac{\kappa^2 \zeta^2}{2} \right] } 
 \exp{ \left[ - {m_{f}^2 \bar{x} + m_{\bar{f}}^2 x \over 2 \kappa^2 x \bar{x} } \right]}
\label{pion-hwf-quark-masses}
\end{equation}
where $\mathcal{N}$ is a normalization constant which is fixed by requiring that 
 \begin{equation}
 	\int \mathrm{d}^2 \mathbf{b} \mathrm{d} x |\Psi(x,\zeta^2)|^2 = 1 \;. 
 	\label{norm}
 \end{equation}
To make predictions, we use constituent quark masses: $[m_q,m_s]=([330,500] \pm 30)$ MeV. Previous work \cite{Forshaw:2012im,Brodsky:2014yha,Brodsky:2016rvj,Deur:2016opc,Ahmady:2016ujw} hints towards a universal AdS/QCD scale: $\kappa \sim 500$ MeV. Here we use $\kappa=523 \pm 24$ MeV \cite{Deur:2016opc}. 

\section{Dynamical spin effects}
\label{Spin}
To account for dynamical spin effects, we assume that 
\begin{equation}
	\Psi^{P}_{h \bar{h}}(x,\mathbf{k}) = \Psi(x, \mathbf{k}) S^P_{h \bar{h}} (x, \mathbf{k})
	\label{spin-improved-wf}
	\end{equation}
where 
\begin{equation}
	S^P_{h \bar{h}} (x, \mathbf{k})= \frac{\bar{u}_{h}(x,\mathbf{k})}{\sqrt{\bar{x}}} \left[A \frac{M_P}{2P^+} \gamma^+ \gamma^5 + B  \gamma^5 \right] \frac{v_{\bar{h}}(x,\mathbf{k})}{\sqrt{x}} 
	\label{spin-structure} 
	\end{equation}
and $\Psi(x, \mathbf{k})$ is the holographic meson light-front wavefunction. $A$ and $B$ are dimensionless, arbitrary constants. 
It therefore follows that \cite{Ahmady:2018muv}
	\begin{eqnarray}
	 	\Psi_{h,\bar{h}}^{P}(x,\mathbf{k})= \left[ (AM_P x\bar{x} + B (xm_f + \bar{x} m_{\bar{f}}) ) h\delta_{h,-\bar{h}}  - B    k e^{-ih\theta_k}\delta_{h,\bar{h}}	\right] \frac{\Psi (x,\mathbf{k}^2)}{x\bar{x}} \;.
	 \label{spin-improved-wfn-k}
	 \end{eqnarray}
	After a two-dimensional Fourier transform of Eq. \eqref{spin-improved-wfn-k}, we obtain
	 \begin{eqnarray}
	 	\Psi_{h,\bar{h}}^{P}(x,\mathbf{b})= \left[ (AM_P x\bar{x} + B(xm_f + \bar{x} m_{\bar{f}})  ) h\delta_{h,-\bar{h}} + i B h   \partial_b \delta_{h,\bar{h}}	\right] \frac{\Psi (x,\zeta^2)}{x\bar{x}}
	 	\label{spin-improved-wfn-b}
	 \end{eqnarray}
 where $\Psi(x,\zeta^2)$ is the two dimensional Fourier transform of the holographic meson wavefunction given by Eq. \eqref{pion-hwf-quark-masses}. With $A=\kappa/\sqrt{2 \pi} M_P$ and $B=0$, we recover the normalized original holographic wavefunction, Eq. \eqref{pion-hwf-quark-masses}:
 \begin{equation}
 	\Psi_{h\bar{h}}^{P} (x,\zeta^2) = \frac{\kappa}{\sqrt{\pi}} \sqrt{x \bar{x}}  \exp{ \left[ -{ \kappa^2 \zeta^2  \over 2} \right] } \exp{ \left[ - {m_{f}^2 \bar{x} + m_{\bar{f}}^2 x \over 2 \kappa^2 x \bar{x} } \right]} \times \frac{1}{\sqrt{2}} h\delta_{h,-\bar{h}} 
\end{equation}
 with a non-dynamical spin wavefunction. 
 
 Our spin-improved wavefunction is normalized using
	\begin{equation}
 	\int \mathrm{d}^2 \mathbf{b} \mathrm{d} x |\Psi^{P}(x,\mathbf{b}^2)|^2 = 1 
 	\label{norm-spin}
 \end{equation}	
 where 
 \begin{equation}
 	|\Psi^{P}(x,\mathbf{b}^2)|^2 =\sum_{h,\bar{h}} |\Psi_{h,\bar{h}}^{P}(x,\mathbf{b}^2)|^2 	\;.
 	\label{sum-notation}
 	\end{equation}
 This normalization condition means that we ignore higher Fock states and it allows us to set $A=1$. Then, the only remaining free parameter is $B$.

\section{Results}
\begin{figure}[htp]
\begin{center}
\includegraphics[width=0.40\textwidth]{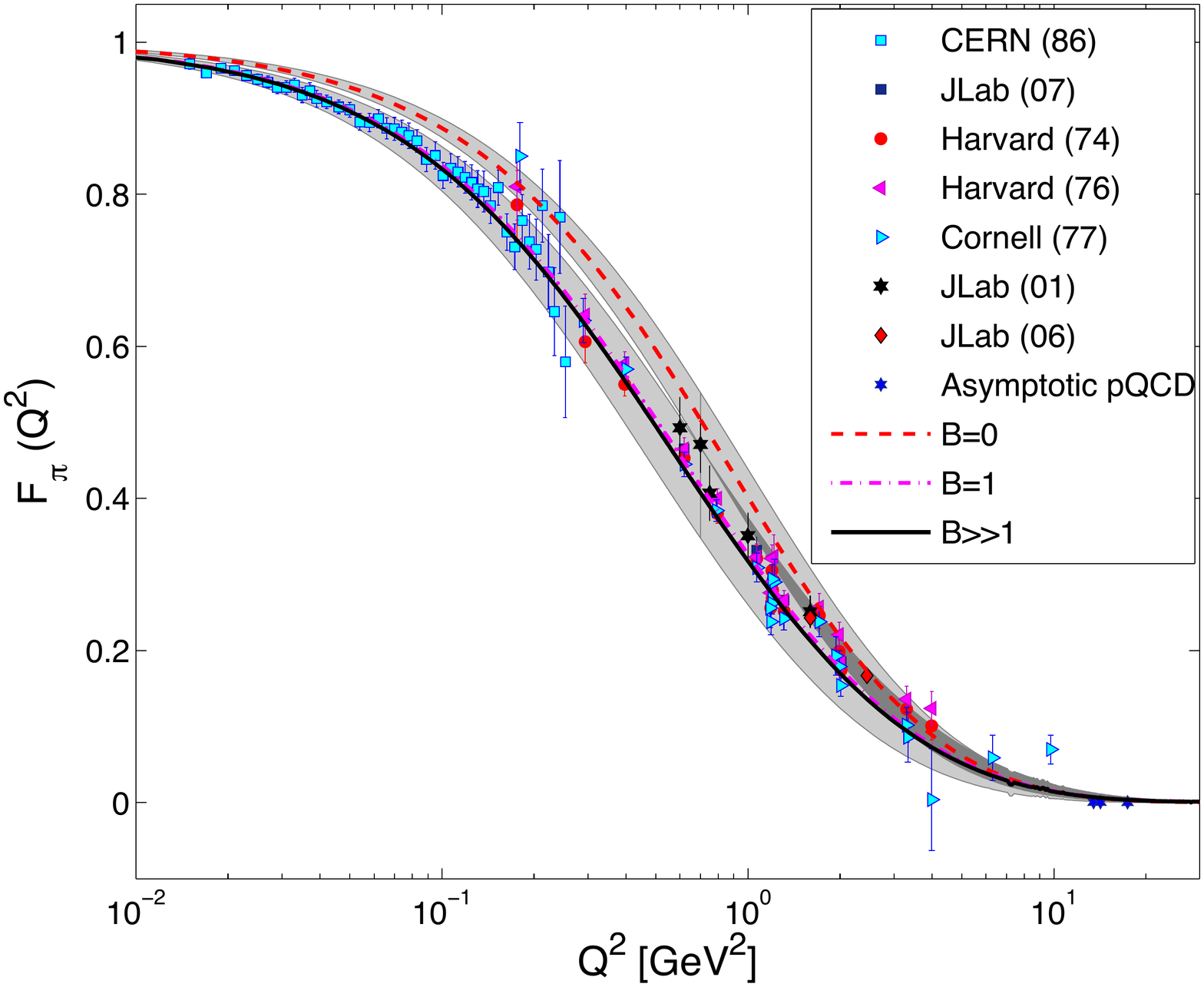}
\includegraphics[width=0.40\textwidth]{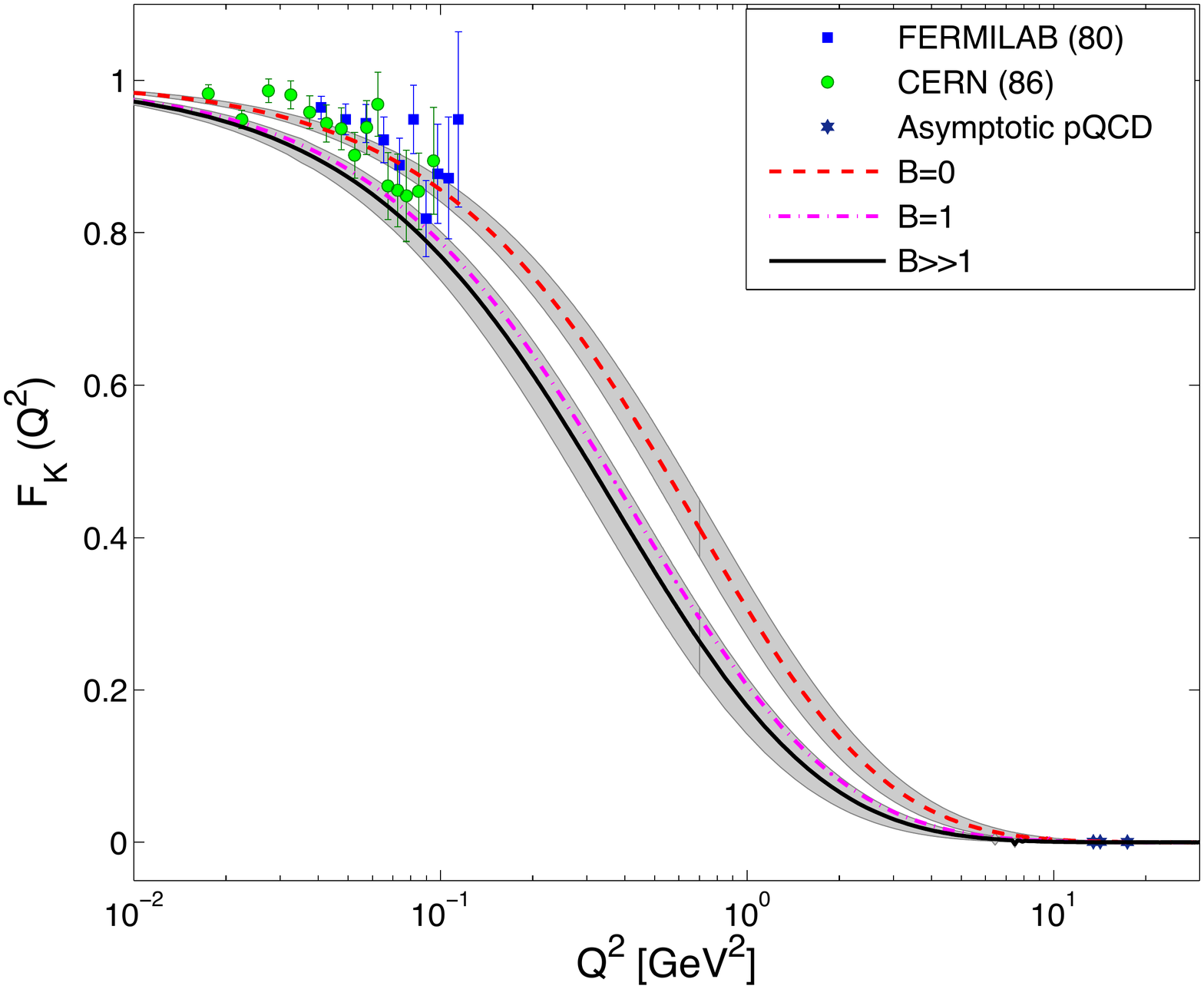}
\caption{(Color online)~Our predictions for the pion (left) and kaon (right) EM form factors. Dashed red curves: $B=0$. Orange dot-dashed curves: $B=1$. Solid black curves: $B \gg 1$. The grey bands for the $B=0$ and $B \gg 1$ curves indicate the theory uncertainty resulting from the uncertainties in the constituent quark masses and the AdS/QCD scale. The references for the data can be found in \cite{Ahmady:2018muv}.
}
\label{Fig:EMFF}
\end{center}
\end{figure}

We compute the decay constants using \cite{Ahmady:2018muv}
\begin{equation}
	f_{P}(m_f,m_{\bar{f}},M_P,B)= 2 \sqrt{\frac{N_c}{\pi}}  \int \mathrm{d} x   [x\bar{x} M_P+ B (x m_f+\bar{x}m_{\bar{f}})]  \left.\frac{\Psi (x,\zeta)}{x\bar{x}}\right|_{\zeta=0}	
\label{decayconstant}
\end{equation}
Our results are shown in Table \ref{tab:decay-constants}. We also compute the root-mean-square pion radius using \cite{Brodsky:2007hb}
\begin{equation}
	\sqrt{\langle r_{P}^2 \rangle} = \left[\frac{3}{2} \int \mathrm{d} x \mathrm{d}^2 \mathbf{b} [b (1-x)]^2 |\Psi^{P}(x,\mathbf{b})|^2 \right]^{1/2} 
	\label{radius}
\end{equation}
where $|\Psi^{P}(x,\mathbf{b})|^2$ is given by Eq. \eqref{sum-notation}, and the EM form factor is given by 
\begin{equation}
	F_{P}(Q^2)= 2 \pi \int \mathrm{d} x \mathrm{d} b ~ b ~ J_{0}[(1-x)  b Q] ~ |\Psi^{P}(x,\textbf{b})|^2 \;.
\label{DYW}
\end{equation}
Our predictions for the charge radii and EM form factors are compared to data in Table \ref{tab:radii} and Figure \ref{Fig:EMFF} respectively. We observe that the pion data favour maximal dynamical spin effects\footnote{This was already reported in \cite{Ahmady:2016ufq}.}, while, on the other hand, the kaon data prefer no dynamical spin effects at all.

\begin{table}
  \centering
  \begin{tabular}{|c|c|c|c|}
    \hline
$P$ &B&$f^{\mathrm{Th.}}_{P}$ [MeV]&$f^{\mathrm{Exp.}}_{P}$[MeV]\\
\hline
 &$0$ & $162 \pm 8$&  \\
\cline{2-3}
$\pi^{\pm}$&$1$ & $138 \pm 5$& $130 \pm 0.04 \pm 0.2$ \\
\cline{2-3}
 &$\gg 1$ & $135 \pm 6$&  \\\hline        
 &$0$ & $156 \pm 8$ &  \\
 \cline{2-3}
 $K^{\pm}$&$1$& $ 142\pm 7$ & $156 \pm 0.5$\\
 \cline{2-3}
 &$\gg 1$& $135 \pm 6$ &  \\
\hline 
         
\end{tabular}
  \caption{Our predictions for the decay constants of the charged pion and kaon, compared to the measured values from PDG \cite{Patrignani:2016xqp}.}
  \label{tab:decay-constants}
\end{table}

\begin{table}
  \centering
  \begin{tabular}{|c|c|c|c|}
    \hline
$P$ &$B$&$\sqrt{\langle r_{P}^2 \rangle}_{\mathrm{Th.}}$ [fm]&$\sqrt{\langle r_{P}^2 \rangle}_{\mathrm{Exp.}}$[fm]\\
\hline
$\pi^{\pm}$ &$0$ & $0.543 \pm 0.022$&  \\
\cline{2-3}
&$1$ & $ 0.673\pm 0.034 $& $0.672 \pm 0.008$ \\
\cline{2-3}
&$\gg 1$ & $0.684 \pm 0.035$&  \\
\hline
$K^{\pm}$ &$0$ & $0.615 \pm 0.038$ & \\
\cline{2-3}
 &$1$& $ 0.778 \pm 0.065$ & $0.560 \pm 0.031$  \\
\cline{2-3}
 &$\gg 1$& $0.815 \pm 0.070$ &  \\
\hline 
         
\end{tabular}
  \caption{Our predictions for the radii of $\pi^{\pm}$ and $K^{\pm}$, compared to the measured values from PDG \cite{Patrignani:2016xqp}.}
  \label{tab:radii}
\end{table}

To make predictions for the $\eta$ and $\eta^\prime$ mesons, we need to account for quantum mechanical mixing. In the SU(3) octet-singlet basis:
\begin{eqnarray}
	|\eta_1\rangle &=&\frac{1}{\sqrt{3}}\left(|u\bar{u}\rangle + |d\bar{d}\rangle + |s\bar{s}\rangle \right)\\ \nonumber
	|\eta_8\rangle &=&\frac{1}{\sqrt{6}}\left(|u\bar{u}\rangle + |d\bar{d}\rangle -2 |s\bar{s}\rangle \right) \;, 
\end{eqnarray}
the physical $\eta$ and $\eta^\prime$ states are given by 
\begin{eqnarray}
\label{mix}	
	|\eta\rangle &=& \cos \theta |\eta_8 \rangle - \sin \theta |\eta_1 \rangle \\ \nonumber
	|\eta^\prime\rangle &=&\sin \theta |\eta_8 \rangle + \cos \theta |\eta_1 \rangle 
\end{eqnarray}
where $\theta$ is the mixing angle. Here we use $\theta=-14.1^\circ \pm 2.8^\circ$ \cite{Christ:2010dd}. Inverting Eq. \eqref{mix} and using the light-front Schr\"odinger Equation, $H_{\mathrm{LF}}|P\rangle = M^2 |P\rangle$ \cite{Brodsky:1997de}, we can express the masses of $\eta_1$ and $\eta_8$ in terms of the physical $\eta$ and $\eta^{\prime}$ masses:
\begin{eqnarray}
M_{\eta_1}^2&=&\sin^2 \theta M_{\eta}^2 + \cos^2 \theta M^2_{\eta^\prime}\\ \nonumber
M_{\eta_8}^2&=&\cos^2 \theta M_{\eta}^2 + \sin^2 \theta M^2_{\eta^\prime}	\;.
\end{eqnarray}
The decay constants of $\pi^0$, $\eta_1$ and $\eta_8$ can then be computed using the axial-vector currents
\begin{equation}
	J^{\pi^0}_{\mu 5}=\frac{1}{\sqrt{2}}(\bar{u} \gamma_\mu \gamma_5 u - \bar{d} \gamma_\mu \gamma_5 d) \;, 
\end{equation}
\begin{equation}
	J^{\eta_8}_{\mu 5}=\frac{1}{\sqrt{6}}(\bar{u} \gamma_\mu \gamma_5 u + \bar{d} \gamma_\mu \gamma_5 d -2 \bar{s} \gamma_\mu \gamma_5 s) \;, 
\end{equation}
and
\begin{equation}
	J^{\eta_1}_{\mu 5}=\frac{1}{\sqrt{3}}(\bar{u} \gamma_\mu \gamma_5 u + \bar{d} \gamma_\mu \gamma_5 d + \bar{s} \gamma_\mu \gamma_5 s)  
\end{equation}
respectively. Assuming isospin symmetry, this leads to 
\begin{equation}
	f_{\pi^0}= f_P(m_{q},M_{\pi^0},B)=f_{\pi^{\pm}} \;,
\end{equation}
\begin{equation}
	f_{\eta_1}= \frac{1}{3}[2f_P(m_{q},M_{\eta_1},B_{q})+ f_P(m_s,M_{\eta_1},B_s)]
\end{equation}
and
\begin{equation}
	f_{\eta_8}= \frac{1}{3}[f_P(m_{q},M_{\eta_8},B_{q})+ f_P(m_s,M_{\eta_8},B_s)] 
\end{equation}
where $f_P(m_{q/s},M_P,B_{q/s})$ is given by Eq. \eqref{decayconstant}. Notice that, based on our findings for the pion and kaon, we have allowed the parameter $B$ to differ in the non-strange and the strange sectors of the $\eta$ and $\eta^\prime$. 

\begin{figure}[htp]
\begin{center}
\includegraphics[width=0.40\textwidth]{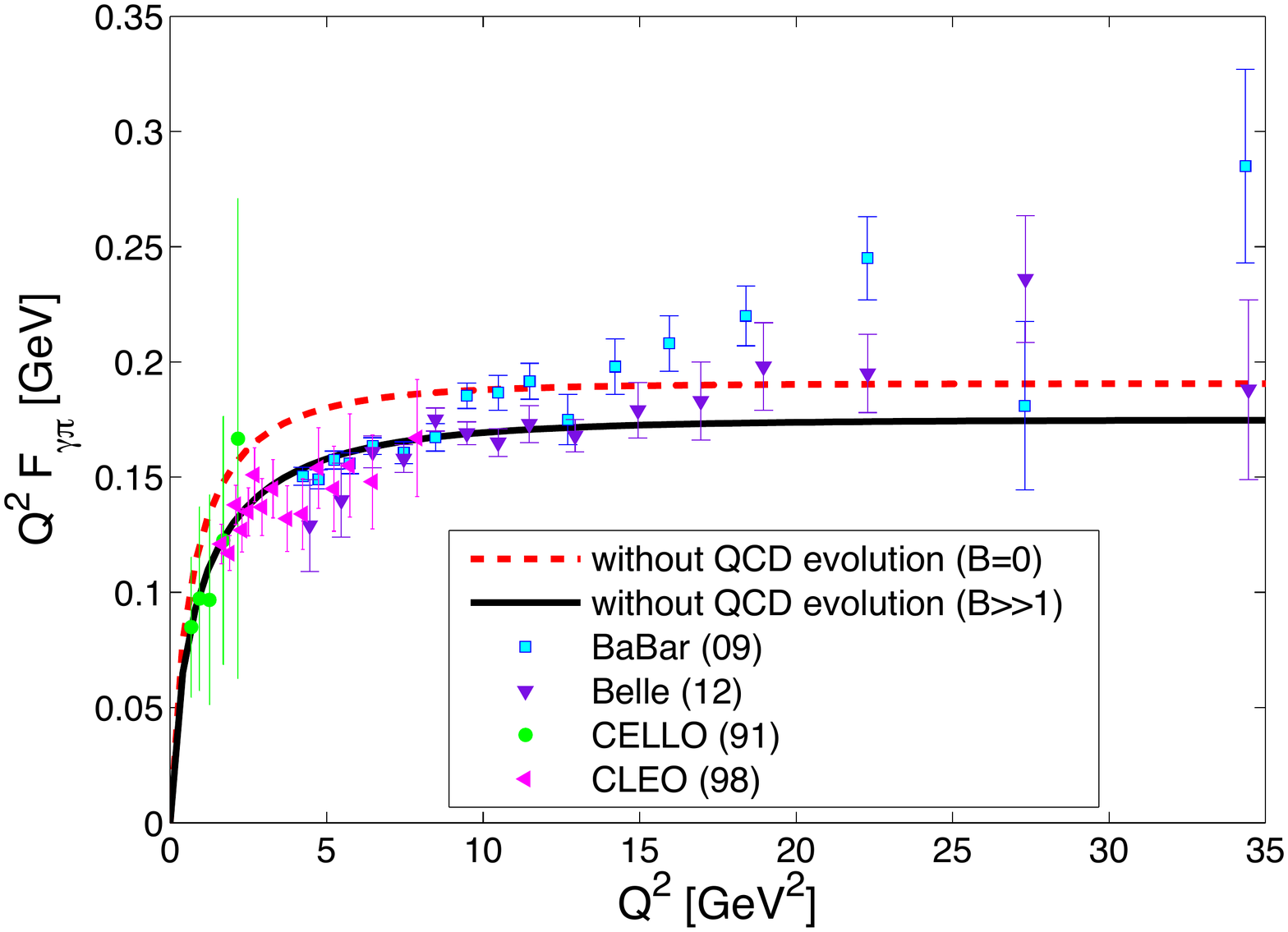}
\includegraphics[width=0.40\textwidth]{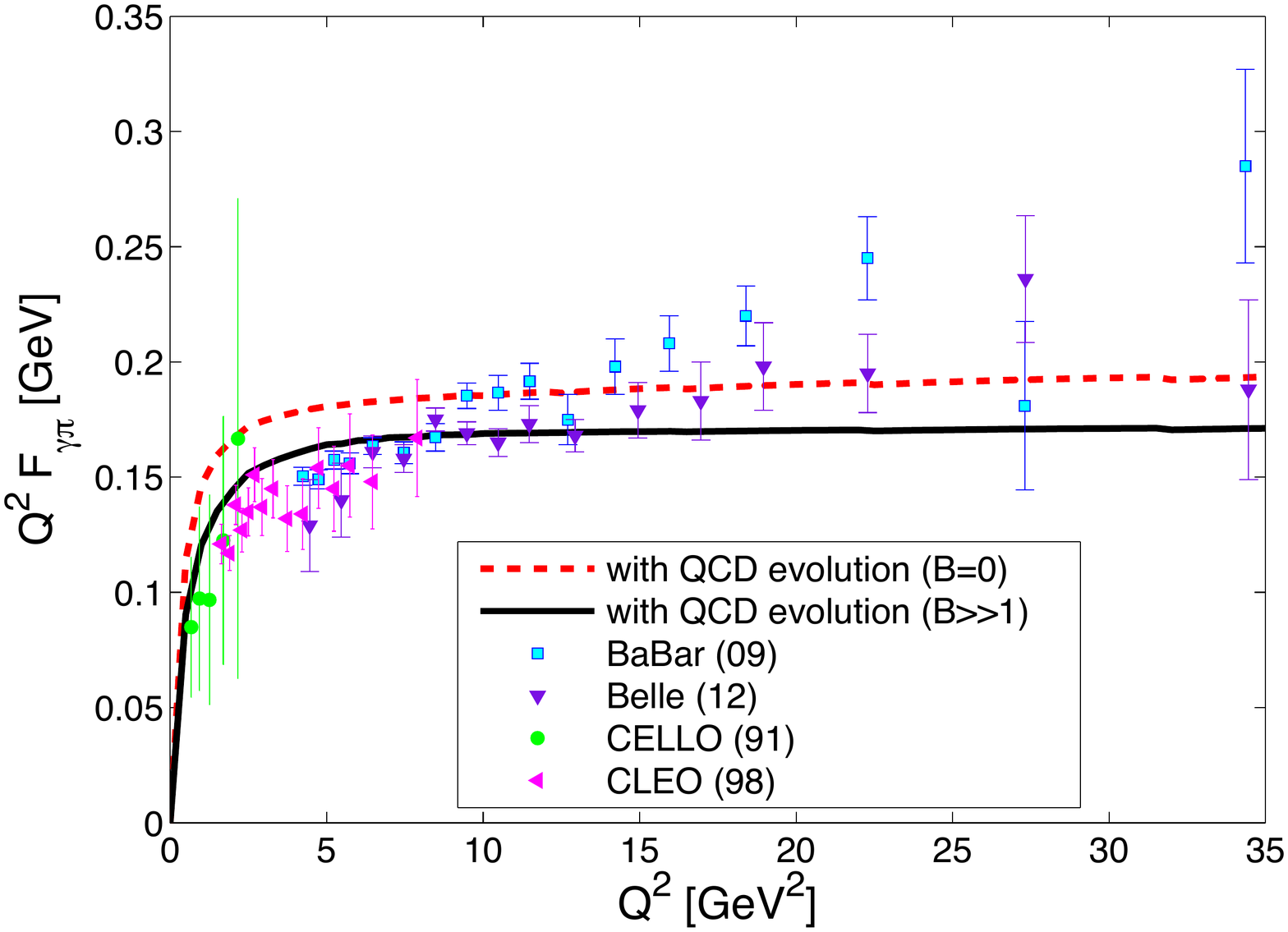}
\caption{(Color online)~Our predictions for the photon-to-pion transition form factor without QCD evolution (left) and with QCD evolution (right). Dashed red curves: $B=0$. Solid black curves: $B \gg 1$. References for the data can be found in \cite{Ahmady:2018muv}.
}
\label{fig:TFF}
\end{center}
\end{figure}

  The photon-meson transition form factor is directly related to the inverse moment of the meson's PDA. For the pion, we have \cite{Choi:2017zxn}
\begin{equation}
	F_{\gamma \pi} (Q^2)= \left(\frac{\hat{e}_{u}^2-\hat{e}_{d}^2}{\sqrt{2}}\right) I(Q^2;m_{q},M_\pi,B)\;,
	\label{TFF}
\end{equation}
where \cite{Lepage:1980fj}
\begin{equation}
	I(Q^2;m_q,M_\pi,B)= 2 \int_0^1 \frac{\varphi_P(x,xQ; m_q,M_\pi,B)}{x Q^2} 
\end{equation}
and $\hat{e}_{u,d}=[2/3,-1/3]$. The PDA, explicitly derived in \cite{Ahmady:2018muv}, is evaluated at a scale $\mu=xQ$ \cite{Brodsky:2011yv}. Our predictions for the pion transistion form factor is shown in Figure \ref{fig:TFF} and they clearly indicate that dynamical spin effects are required to describe the data. 

For the $\eta$ and $\eta^\prime$, taking into account mixing, their transition form factors are given by 
\begin{eqnarray}
	F_{\eta \gamma} &=& \cos \theta F_{\eta_8 \gamma} - \sin \theta F_{\eta_1 \gamma} \\ \nonumber
	F_{\eta^\prime \gamma} &=& \sin \theta F_{\eta_8 \gamma} + \cos \theta F_{\eta_1 \gamma} \\ \nonumber
\end{eqnarray}	
where
\begin{equation}
	F_{\eta_1 \gamma}(Q^2)= \left(\frac{\hat{e}_{u}^2+\hat{e}_{d}^2}{\sqrt{3}}\right) I(Q^2;m_{q},M_{\eta_1},B_{q}) + \frac{\hat{e}_s^2}{\sqrt{3}} I(Q^2;m_s, M_{\eta_1},B_s)
\end{equation}
and
\begin{equation}
	F_{\eta_8\gamma}(Q^2)= \left(\frac{\hat{e}_{u}^2+\hat{e}_{d}^2}{\sqrt{6}}\right) I(Q^2;m_{q},M_{\eta_8},B_{q})-2 \frac{\hat{e}_s^2}{\sqrt{6}} I(Q^2;m_s, M_{\eta_8},B_s) \;.
\end{equation}
We note that, in the SU(3) chiral limit $\{m_q,m_s,M_P\} \to 0$, the transition form factors of the 3 mesons differ only by a constant factor: $F_{\eta_1 \gamma}(Q^2)=(2\sqrt{2}/\sqrt{3}) F_{\pi \gamma}(Q^2)$ and  $F_{\eta_8 \gamma}(Q^2)=(1/\sqrt{3}) F_{\pi \gamma}(Q^2)$.  

\begin{figure}[htp]
\begin{center}
\includegraphics[width=0.40\textwidth]{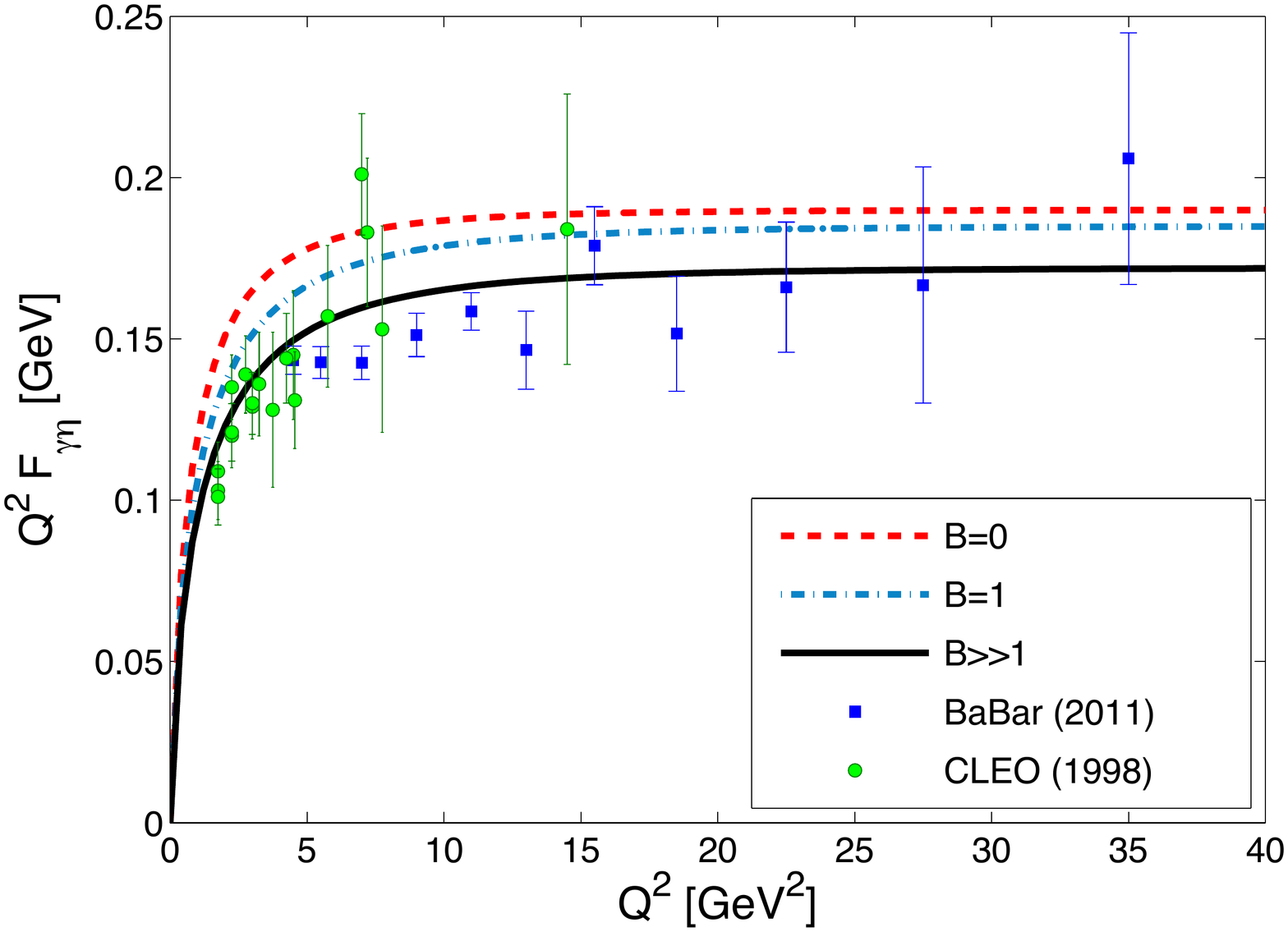}
\includegraphics[width=0.40\textwidth]{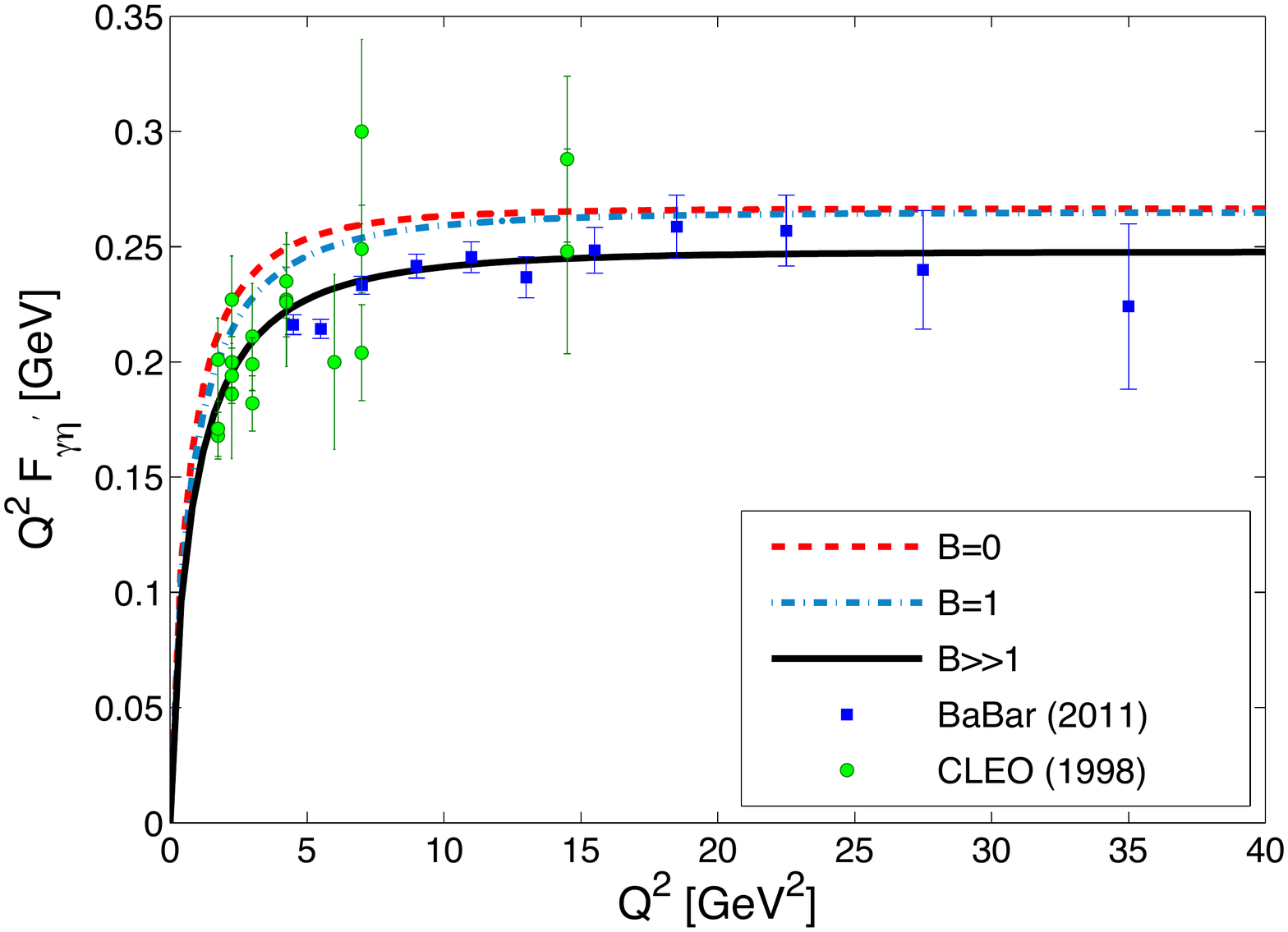}
\caption{(Color online)~Our predictions for the $\eta$ and $\eta^\prime$ transition form factors compared to the data from the BaBar and CLEO  Collaborations. For these predictions, we have set $B_q=B_s=B$. Dashed red curves: $B=0$. Dot-dashed blue curves: $B=1$. Solid black curves: $B \gg 1$. References for the data can be found in \cite{Ahmady:2018muv}.
}
\label{Fig:eta-etaprime-TFF}
\end{center}
\end{figure}

Our predictions are shown in Figure \ref{Fig:eta-etaprime-TFF}. As can be seen, our predictions for the $\eta$ and $\eta^\prime$ agree very well with the data when dynamical spin effects are maximal in both mesons. In \cite{Ahmady:2018muv}, we also predicted the radiative decay width of the $\eta$ and $\eta^\prime$ and found that for the $\eta^\prime$, dynamical spin effects are required to fit the decay width datum while this is not the case for the $\eta$.  A fully analysis of the $\eta-\eta^\prime$ system is required to shed light on this tension but, based on our predictions for the (larger) transition form factor data set, it is likely that dynamical spin effects are also important in the $\eta-\eta^\prime$ system.

\section{Conclusions}
We have investigated the importance of dynamical spin effects in the pion, kaon and $\eta-\eta^\prime$ system. We find that while they are  crucial to describe the pion data, they are not necessary to accommodate the available kaon data. On the other hand, they are required to describe the transition form factor data for the $\eta$ and $\eta^\prime$ mesons. We conclude that the importance of dynamical spin effects is more likely to be correlated with the flavour content of the pseudoscalar mesons than to their masses. 
\section{Acknowledgements}
R.S thanks the organizers of Confinement 2018 for a successful conference and Stan Brodsky for useful discussions. M.A and R.S are supported by individual Discovery Grants from the National Science and Engineering Research Council of Canada (NSERC): SAPIN-2017-00033 and SAPIN-2017-00031 respectively. C.M is supported by the China Postdoctoral Science Foundation (CPSF) under the Grant No. 2017M623279 and the National Natural Science Foundation of China (NSFC) under the Grant No. 11850410436.
\bibliographystyle{spphys}       
\bibliography{skeleton.bib}   

\end{document}